\documentclass{emulateapj}
\usepackage{epsfig}
\submitted{{\it Accepted for publication in AJ}}

\newcommand\dgr{$\delta_{(g-r)}$}
\newcommand\kms{km s$^{-1}$}
\newcommand\feh{$[$Fe$/$H$]$}

\defcitealias{west10}{Paper I}

\shortauthors{Bochanski et al.}
\shorttitle{Statistical Parallax of SDSS M Subdwarfs}
\begin{document}

\title{Mapping the Local Halo: Statistical Parallax Analysis of SDSS
  Low--Mass Subdwarfs}

\slugcomment{Accepted by AJ}

\author{		John J. Bochanski\altaffilmark{1,2},	
  Antonia Savcheva\altaffilmark{3,4}, 
  		 Andrew A. West\altaffilmark{3},
 Suzanne L. Hawley\altaffilmark{5},
	 }
\altaffiltext{1}{Haverford College, 370 Lancaster Ave, Haverford PA
  19041 USA\\
email:jbochans@haverford.edu}
\altaffiltext{2}{Astronomy and Astrophysics Department, Pennsylvania
  State University, 525 Davey Laboratory, University Park, PA 16802 USA }

\altaffiltext{3}{Department of Astronomy, Boston University, 725 Commonwealth Avenue, Boston, MA 02215 USA}
\altaffiltext{4}{Harvard-Smithsonian Center for Astrophysics, 60
  Garden st., Cambridge, MA 02138 USA}
\altaffiltext{5}{Astronomy Department, University of Washington,
   Box 351580, Seattle, WA  98195 USA}

\begin{abstract}
We present a statistical parallax study of nearly 2,000 M
subdwarfs with photometry and spectroscopy from the Sloan Digital Sky
Survey. Statistical parallax analysis yields the
mean absolute magnitudes, mean velocities and velocity ellipsoids for
homogenous samples of stars.   We selected homogeneous groups of
subdwarfs based on their photometric colors and spectral appearance.
We examined the
color--magnitude relations of low--mass subdwarfs and quantified their
dependence on the newly-refined
metallicity parameter, $\zeta$.   We also developed a photometric metallicity
parameter, {\dgr}, based on the $g-r$ and $r-z$ colors of low--mass
stars and used it to select stars with similar metallicities.  The kinematics of low--mass
subdwarfs as a function of color and metallicity were also examined
and compared to main sequence M dwarfs. We find that the SDSS
subdwarfs share similar kinematics to the inner halo and thick disk.
The color-magnitude
relations derived in this analysis will be a powerful tool for
identifying and characterizing low--mass metal--poor subdwarfs in future surveys
such as GAIA and LSST, making them important and plentiful tracers of the stellar halo.
\end{abstract}

\section{Introduction}
Modern surveys, such as the Sloan Digital Sky Survey
\citep[SDSS;][]{2000AJ....120.1579Y} and the Two Micron All--Sky Survey
\citep[2MASS;][]{2006AJ....131.1163S}, have provided a detailed picture
of the Milky Way's structure, kinematics and evolution.  The Galaxy's
disks and halo have been measured by a slew of stellar tracers,
including RR Lyrae stars, blue horizontal branch stars and main--sequence turnoff stars
\citep[i.e.,][]{2001ApJ...554L..33V,2008ApJ...680..295B,2009ApJ...700.1282Y}.  But these tracers only represent a small fraction of the
stars in the Milky Way.  The most dominant stellar inhabitant of the Milky Way, in
mass and number, are the low--mass M
dwarfs \citep{2002AJ....124.2721R, 2003PASP..115..763C,2010AJ....139.2679B}.
Recent surveys have observed these stars in large numbers
with deep, precise multi-band photometry over large fractions of the
sky.  The resulting photometric and spectroscopic databases are the
largest ever assembled for M dwarfs \citep{2010AJ....139.2679B, 2011AJ....141...97W} and have
led to novel investigations of the Galaxy.  Most notably, 
the M dwarf stellar density distributions \citep{2010AJ....139.2679B, 2008ApJ...673..864J} and kinematic
structures \citep{2007AJ....134.2418B,2010ApJ...716....1B,pineda} have been used as tracers
of the Galaxy's gravitational potential.  

The oldest members of the low--mass stellar populations are the
subdwarfs.  First coined by \cite{1939ApJ....89..548K}, subdwarfs are
low--metallicity stars ({\feh} $\lesssim -1$) from earlier generations of
star formation.   The decreased opacity from metals leads to smaller
stellar radii at the same mass, as first noted by
\cite{1959MNRAS.119..278S}.  In observational terms, low--metallicity
subdwarfs lie below the main sequence of stars, with a fainter
absolute magnitude at the same color.  Detailed studies of the low--mass subdwarfs are
hindered by their low luminosities, but observational efforts have
identified low--mass subdwarfs throughout the M and L spectral
sequence \citep{1997AJ....113..806G, 2003ApJ...592.1186B,
  2003MNRAS.344..583D, 2012A&A...542A.105L}.  Subdwarfs are relatively
rare in the solar neighborhood, with only 3 known within 10 pc, compared to 243 main sequence
systems \citep{2006ApJ...638..446M}.  Traditionally, they have been
identified by their large proper motions, such as Kapetyn's star
\citep[$\mu = 8.66 ^{\prime\prime} /$
yr$^{-1}$;][]{1976ApJ...210..402M,2004MNRAS.350..575W,2007A&A...474..653V}.
Despite their paucity in the solar neighborhood, they probably comprise
the majority of stars in the Milky Way's stellar halo.
Precise metallicities for most M subdwarfs have not been measured, due
to the lack of adequate stellar models \citep{2011ASPC..448..531W}.  Instead, considerable effort
has gone into developing proxies for metallicity, most notably
$\zeta$.  Introduced by \cite{2007ApJ...669.1235L} and revised by \cite{2012AJ....143...67D},
$\zeta$ is based on the molecular bandhead indices originally employed
by \cite{1997AJ....113..806G} and crudely
traces metallicity \citep{2009PASP..121..117W}.  While $\zeta$ does
well at estimating {\feh} at low metallicity, it has limited
precision near solar values.   New empirical calibrations in the
infrared and optical \citep{2012ApJ...748...93R,2012arXiv1211.4630M} may improve the
calibration between $\zeta$ and an absolute metallicity scale ({\feh}). 

Despite the advances in survey technology, precise 
distances of SDSS low--mass dwarfs remain elusive.  Most
color--absolute magnitude relations (CMRs) used to obtain photometric
parallax distances, 
are populated by $\sim~100$ nearby stars, biasing the metallicity
distribution to values similar to that of the Sun \citep{1992AJ....103..638M, 1995AJ....110.1838R, bochanskithesis,
  2011AJ....141..117J}.  This is particularly troubling for calibrating the SDSS CMRs, with low--mass
dwarfs detected out to a few kpc from the Sun.  Complementary investigations of
solar--mass stars in SDSS
\citep{Ivezic08, 2010ApJ...716....1B} have shown a gradient in metallicity with position
in the Galaxy, as well as a shift in the color--absolute magnitude
locus \citep{2008ApJS..179..326A}.  Using photometry and trigonometric
parallaxes of nearby stars, 
\cite{2010AJ....139.2679B} derived color--absolute magnitude relations
for M dwarfs in the $ugriz$ filter set.  However, nearly all of the
stars used for the \cite{2010AJ....139.2679B} relations are within 20 pc and have spectra consistent with
solar--metallicity M dwarfs, limiting their applicability to
low--metallicity subdwarfs.

Using a statistical parallax technique, \cite{2011AJ....141...98B} estimated
the absolute magnitudes for SDSS M dwarfs in the Data Release 7
catalog \citep[DR7;][]{2009ApJS..182..543A, 2011AJ....141...97W}.
They demonstrated that nearby low--mass stars
are brighter in $M_r$ than most  early-type M dwarfs in SDSS.  This
was attributed to the influence of metallicity and magnetic
activity on the brightness of SDSS M dwarfs.   However, the metallicity range
probed in the sample was relatively limited, since the \cite{2011AJ....141...97W}
sample mostly contained solar--metallicity M dwarfs.

In this paper, we present a statistical parallax analysis of over 2000
low--mass
subdwarfs identified in the SDSS DR7 catalog.  A summary of the
observations employed in our analysis is contained in Section \ref{sec:obs}.  The
statistical parallax method, which yields absolute magnitudes and
kinematics for a homogeneous groups of stars, is discussed in Section
\ref{sec:method}.  We find that subdwarfs with similar metallicity,
traced by either spectroscopic or photometric indicators, form
distinct color--magnitude relations, well--separated from the solar
metallicity CMRs previously derived.  We quantify these relations and
compare subdwarf kinematics to main--sequence M dwarf kinematics in Section \ref{sec:results},
followed by our conclusions in Section \ref{sec:conclusions}.

\section{Observations}\label{sec:obs}
The statistical parallax method relies on accurate and precise
astrometry, photometry, proper motions and radial velocities.  All of
these quantities are available within SDSS catalogs. Multi-band photometry was collected
using the 2.5m SDSS telescope at Apache Point
Observatory \citep{2006AJ....131.2332G}.  The telescope employed a drift-scan
technique, simultaneously imaging the sky in $ugriz$ along 5 camera
columns \citep{1998AJ....116.3040G}.  Typical faint limits for the $53.9$ second
exposures were $r \sim$ 22 mag, with bright objects saturating
near $r \sim 15$ mag.  Typical systematic photometric precision is 0.02 mag
\citep{2007AJ....134..973I} and has been quantified using repeat scans of ``Stripe
82'', a 300 sq. deg. equatorial patch of sky imaged repeatedly over
the survey lifetime.  

Proper motions were supplied from the \textsc{ProperMotions} table
contained in the Catalog Archive Server \citep[CAS;][]{2005cs........2072O}.  Proper
motions were derived by matching SDSS detections to POSS observations,
with a typical temporal baseline of $\sim$ 50 years \citep{2004AJ....127.3034M}.  This led to
a precision of $\sim 2$ mas yr$^{-1}$.  Several cuts on precision
and error flags were employed to ensure high quality proper motions.
These are described in detail in our previous investigations
\citep[i.e.,][]{2010AJ....139.2566D, 2011AJ....141...98B, 2011AJ....141...97W}.  Astrometric positions are reported by the
SDSS pipeline with an internal precision of 25 mas and an absolute
accuracy of 45 mas in each coordinate direction \citep{2003AJ....125.1559P}.

When the sky conditions at Apache Point were not photometric, the SDSS
telescope obtained $R \sim 1,800$ spectra of objects chosen by
targeting algorithms designed to select high priority objects.  This
includes high--redshift quasars \citep{2009ApJS..180...67R, 2010AJ....139.2360S}, luminous
red galaxies \citep{2001AJ....122.2267E} and other exotic targets, such as
cataclysmic variables \citep{2011AJ....142..181S}.  The fiber-fed spectrographs
\citep{1999AAS...195.8701U} acquired 640 spectra simultaneously.  Over 600,000
stellar spectra have been identified in the latest SDSS Data Release
\citep[DR9; ][]{2012arXiv1207.7137S}.   We measured H$\alpha$ emission and
molecular bandhead strengths using the \textsc{Hammer} pipeline
\citep{2007AJ....134.2398C} on the SDSS subdwarf spectra. Radial velocities, derived from comparing to the SDSS M
dwarf templates \citep{2007AJ....133..531B} were measured for all objects in
the \cite{2011AJ....141...97W} catalog, with a precision of $\sim$ 7 km s$^{-1}$.

The subdwarfs in this analysis passed the initial color-cuts used to
construct the \cite{2011AJ....141...97W} catalog.  Subdwarfs were manually excluded from
that catalog and placed into a separate list for further examination.
There were 6600 spectra that were classified as ``Odd''
during the original spectral typing.  These spectra were re-examined, and 260
low--mass subdwarfs were identified by eye.  We supplemented this list
by selecting 1767 objects in the \cite{2011AJ....141...97W} catalog with $\zeta < $ 0.825
\citep{Lepine03, 2012AJ....143...67D}.   Further details and analysis of this
subdwarf catalog can be found in \cite{antonia}.

\section{Method}\label{sec:method}
The maximum likelihood application of the statistical parallax
techniques employed here was first described in
\cite{1983veas.book.....M} and was applied to SDSS M dwarfs in
\cite{2011AJ....141...98B}.  It has also been used for previous
studies of RR Lyrae stars 
\citep{1986ApJ...302..626H,1986MNRAS.220..413S,1996AJ....112.2110L,1998A&A...330..515F,2012arXiv1208.2689K},
Cepheids \citep{1991ApJ...378..708W} and nearby M dwarfs \citep{1996AJ....112.2799H}.
We refer the reader to \cite{2011AJ....141...98B} for detailed description
of the method and briefly describe it here.  The statistical parallax
method models the kinematic properties of a homogeneous population of
stars with
nine parameters, determining the mean velocities, dispersions and the orientation of the velocity
ellipsoid.  Two additional parameters model the value and
dispersion of the absolute magnitude of the population.  
The eleven parameters used in the model are given in Table
\ref{table:parameters}.  In particular, the distance scale and
absolute magnitude values are given by the
following equation from \cite{1986ApJ...302..626H}:
\begin{equation}
{\sigma_M}^2 = \log_{10}[1 + {\sigma_k}^2/(1 + k)^2]/(0.04~\ln10),
\label{eqn:sigk}
\end{equation}
where $\sigma_M$ is the spread in absolute magnitude for a given
luminosity bin, and $\sigma_k$ is the spread in $k$ which is the distance
scale parameter.  The absolute magnitude is related to $k$ through the following equation:
\begin{equation}
{M} = 5\log_{10}(1 + k) + M_A - 0.1~\ln10 (\sigma_M)^2,
\end{equation}
where $M_A$ is an initial estimate of the absolute magnitude.  

The positions, proper motions, radial velocities and apparent
magnitudes, after correcting for Galactic extinction using the dust
maps of \cite{1998ApJ...500..525S}, are
input for each sample.  The median $r$-band extinction for the entire
sample was 0.07 mag,
with 96\% of the sample having $A_r < 0.4$.  
While most of our stars (99\%) are more than 15 degrees away from the Plane,
we compared $A_r$ for stars closer to the Plane to the reddening maps of
\cite{2011AJ....142...44J}.  The median difference between our adopted
reddening and the \cite{2011AJ....142...44J} maps for these stars was 0.06 mag, well
within the systematic uncertainties of this analysis.
Observational uncertainties are used to properly weight the
solution.  
The eleven parameters in the model
were solved simultaneously by maximizing the likelihood with a
geometric simplex optimization
\citep{optimizationsimplex_nelder_1965,1978Daniels}, while the
uncertainties were estimated using computed derivatives.
The maximum likelihood equations and 
simplex method are described in detail in \cite{1986ApJ...302..626H}.
As in \cite{2011AJ....141...98B}, we fixed $\sigma_k$ at four
values: 0.05, 0.1, 0.2 and 0.3.  We adopt 
$\sigma_k = 0.2$ as a fiducial value, which corresponds to $\sigma_M \sim 0.4$ which is the
typical dispersion in absolute magnitude for low--mass stars in the
solar neighborhood \citep{2010AJ....139.2679B}.

\begin{center}
\begin{deluxetable}{lrr}
\tablewidth{0pt}
\tabletypesize{\small}
 \tablecaption{Statistical Parallax Fit Parameters}
 \tablehead{
 \colhead{Parameter} &
 \colhead{Units} &
 \colhead{Description} 
}
 \startdata
$\sigma_{U}$ & km s$^{-1}$ & Velocity dispersion in radial direction\\
$\sigma_{V}$ & km s$^{-1}$ & Velocity dispersion in orbital direction\\
$\sigma_{W}$ & km s$^{-1}$ & Velocity dispersion in vertical direction\\
\bf{$r,\phi,z$} & radians & Orientation of velocity ellipsoid\\
$U$ & km s$^{-1}$ & Solar peculiar motion ($r$)\\
$V$  & km s$^{-1}$& Solar peculiar motion ($\phi$)\\
$W$ &km s$^{-1}$ & Solar peculiar motion ($z$)\\
$k$ & \nodata & Distance scale\\
$\sigma_k$ & \nodata & Dispersion in distance scale
\enddata
 \label{table:parameters}
\end{deluxetable}
\end{center}

\subsection{Constructing Subsamples for Analysis}
\begin{figure*}[htbp]
  \centering
  \includegraphics[scale=0.5, angle=90]{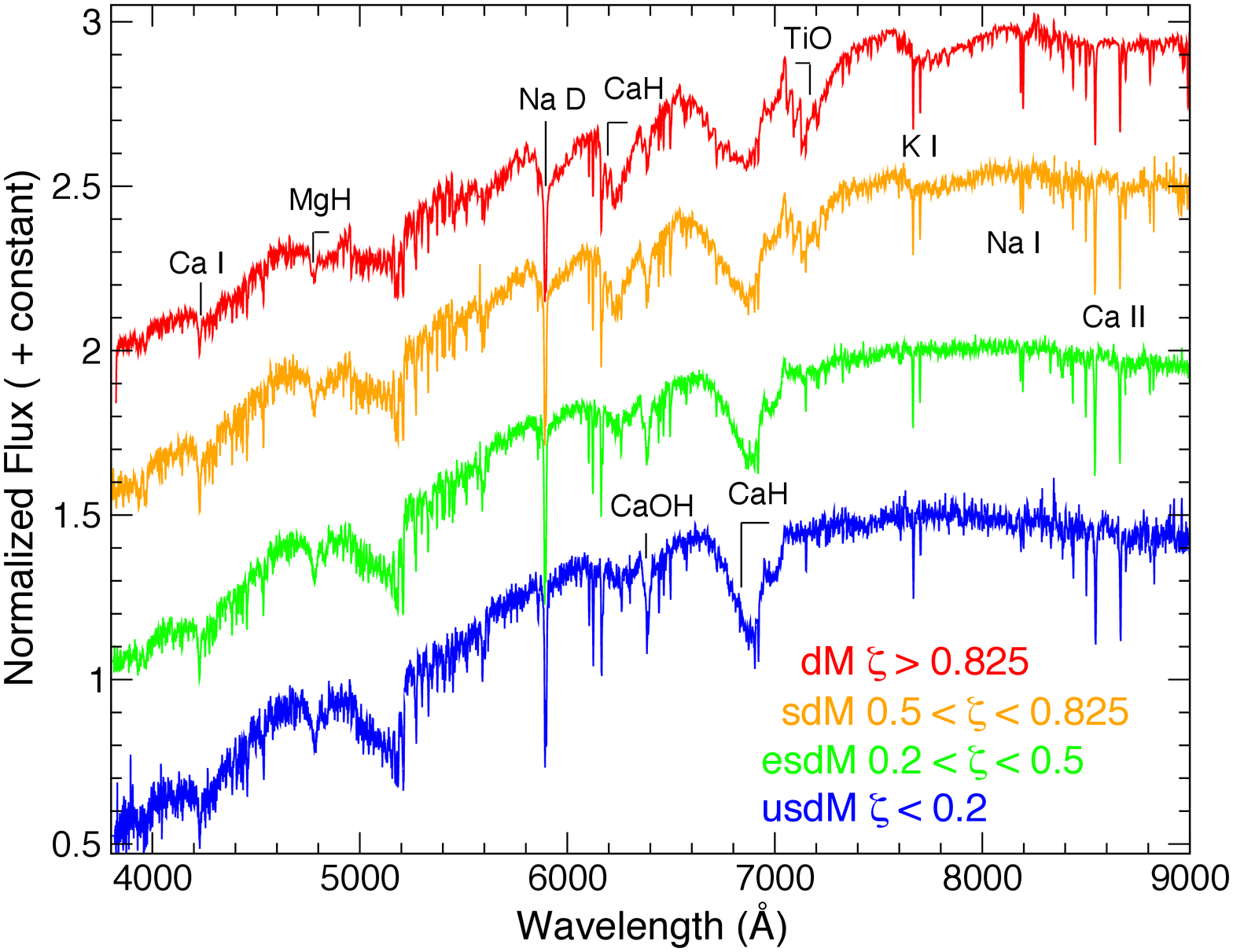}
\caption{Example spectra of a solar--metallicity M dwarf (dM-red), a
  subdwarf (sdM-orange), extreme subdwarf (esdM-green) and ultra--subdwarf (usdM-blue).
  The boundaries between the \cite{2007ApJ...669.1235L} metallicity subclasses are
  given in the legend.
  Major atomic and molecular spectral features are marked.  
  The CaH near 6800 \AA\ grows stronger with
  decreasing metallicity, while TiO near 7100 \AA\ weakens.  The Na I
  doublet near 8200 \AA\ also weakens with declining metallicity.
}
  \label{fig:spec}
\end{figure*}

Statistical parallax analysis requires the selection of a homogenous
set of stars.  Each set must be moderately populated (i.e., $N \gtrsim 30$) to
ensure a well--measured velocity ellipsoid.  We performed a
series of cuts on our subdwarf sample to explore the photometric and
kinematic properties of SDSS low--mass subdwarfs.   They are listed in Table
\ref{table:subsamples} and described here.  
First, we separated the subdwarfs by $r-z$ color.  This color has been
shown to track $M_r$ well \citep{2010AJ....139.2679B}.  However,
metallicity can also influence the
color--absolute magnitude loci.  Thus, we further divided the sample by
$r-z$ color and subdwarf subclass. The subdwarfs (sdM), extreme
subdwarfs (esdM) and ultra--subdwarfs (usdM), were chosen using the
definitions of \cite{2007ApJ...669.1235L} and \cite{2012AJ....143...67D}.  Example spectra of
the different metallicity subclasses and their correspondence to
$\zeta$ are shown in Figure
\ref{fig:spec}.  The classes are separated by $\zeta$, which is defined by the CaH
and TiO optical bandhead strengths.   Past studies have employed the TiO bandheads as tracers of absolute magnitude
\citep[i.e., ][]{1995AJ....110.1838R}, while combinations of CaH and TiO have been used
to trace metallicity shifts \citep{1997AJ....113..806G, 2006ApJ...645.1485B, 2007ApJ...669.1235L, 2007AJ....134.2418B}.  
As seen in Figure \ref{fig:spec}, both CaH and TiO display sensitivity
to metallicity difference, with CaH near 6200 \AA\ and 6800 \AA\ strengthening with decreasing
metallicity, while the TiO bandheads weaken. 
 Since low--metallicity subdwarfs trace an older population
\citep{2006ApJ...638..446M}, there is a considerable dearth
of chromospheric H$\alpha$ emission.   Only one color bin ($r-z =
1.65$) contained enough active stars for analysis.  Analysis of this
bin, as described below, followed the general trend observed by
\cite{2011AJ....141...98B}, with active stars being brighter at the same color.  

\begin{figure}[htbp]
  \centering
  \includegraphics[scale=0.4]{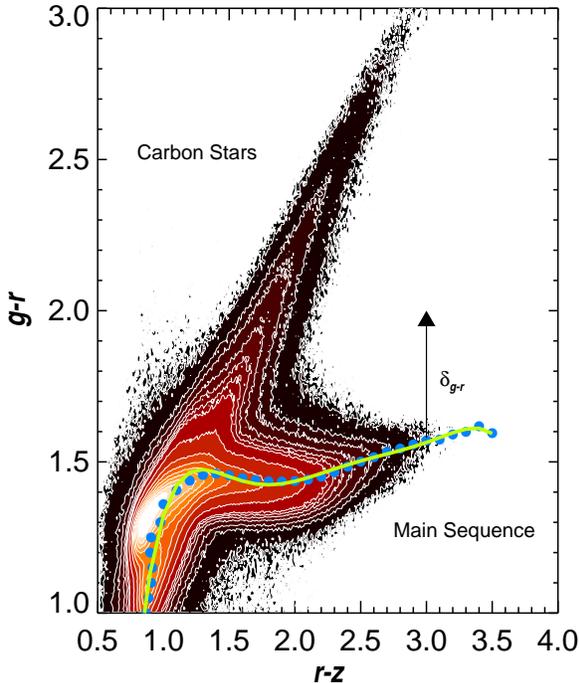}
\caption{The $g-r, r-z$ color-color locus of SDSS DR9 stars.
  The stars with clean and precise photometry were chosen, and the
  main sequence was selected by selecting stars with $g-r < 1.8$.  The $(g-r)_{dM}$ fit
  to the median $g-r, r-z$ colors (blue circles) given in Equation \ref{eqn:gr_dm} is over-plotted in green.  Increasing
  values of {\dgr} are denoted with the arrow.  Low--mass subdwarfs
  generally have redder $g-r$ colors at the same $r-z$ color (or
  spectral type) and would be found above the green line.  This locus
  also traces the $\zeta \sim 1$ bins in Figure 10 of
  \cite{2011AJ....141...97W}. to $\lesssim 0.05$ mag.}
  \label{fig:gr_dm}
\end{figure}

While dividing low--mass subdwarfs into metallicity classes is
useful for spectroscopic investigations, many current and planned
large surveys, such as the Panoramic Survey Telescope \& Rapid Response
System \citep[PanSTARRS;][]{2002SPIE.4836..154K} and Large Synoptic Survey
Telescope \citep[LSST;][]{2008arXiv0805.2366I} will consist solely of photometry.
\cite{1977ApJ...214..778H} originally noted that lower metallicity
led to increased hydride absorption in the $B$ band giving a redder $B-V$
color for subdwarfs. \cite{2004AJ....128..426W} demonstrated that the
SDSS $g-r$ color also reddens with
decreasing metallicity. \cite{2011AJ....141...97W} further quantified this effect by fitting a two--dimensional relation between $\zeta$, $g-r$
and $r-z$ (their Equation 3), but their relations are limited to
solar--metallicity M dwarfs.  To quantify the shift in $g-r$ for
subdwarfs, we introduce {\dgr}, a quantity
that measures the difference in $g-r$ between a subdwarf and its
solar--metallcity counterpart as a function of $r-z$ color.  It is
defined as:
\begin{equation}
\delta_{(g-r)} = (g-r)_{subdwarf} - (g-r)_{dM},
\label{eqn:dgr}
\end{equation}
where $(g-r)_{dM}$ is:
\begin{equation}
(g-r)_{dM} =  \sum\limits_{i=0}^6 C_n \times (r-z)^n,  
\label{eqn:gr_dm}
\end{equation}
with the polynomial coefficients ${C_0...C_6}$ given in Table
\ref{table:dgr_def} and $(g-r)_{subdwarf}$ is the $g-r$ color of the subdwarf.  The polynomial fit described in Equation
\ref{eqn:gr_dm} is shown in Figure \ref{fig:gr_dm} and was fit to the
$g-r, r-z$ color-color locus of the main sequence M dwarfs from SDSS DR9.  These
main sequence stars were selected to have good $griz$ photometry
(i.e., $15 < griz < 20$, $\sigma_{griz} < 0.05$, and \textsc{Clean} =
1).  Carbon stars were removed by enforcing $g-r < 1.8$ for the fit
\citep{2002AJ....124.1651M, 2007AJ....134.2398C}.
Median $g-r$ and $r-z$ colors were computed, and a sixth order
polynomial was fit from $0.8 < r-z < 3.6$.

The {\dgr} metric is a crude measure of metallicity, and as discussed below,
traces separate loci in color--magnitude space.   Using the
statistical parallax calibrations discussed below, we will show that with $grz$
photometry alone, one can estimate the metal content and distance of a
low--mass star.  Similar relations in other filter sets have been derived for
nearby stars \citep{2012AJ....143..111J}.  

Finally, we removed RV outliers that
were found more than 3 standard deviations away from the median RV of
each subsample.   Typically, this removed only one or two stars with large RVs
( $\gtrsim 300$ {\kms}).  These were removed to prevent them from
significantly influencing the mean velocity measured by the
statistical parallax routine, since each RV was assumed to have a
fixed uncertainty.  Further details on high velocity subdwarfs are
discussed in \cite{antonia}.

\begin{center}
\begin{deluxetable}{llll}
\tablewidth{0pt}
\tabletypesize{\small} 
\tablecaption{SDSS Subsamples}
 \tablehead{
 \colhead{Initial Cut} &
 \colhead{Bin Size} &
 \colhead{Further Criteria} &
 \colhead{Comments} 
}
 \startdata
$r-z$          &   0.3 mag bins & \nodata & \nodata \\
                 &    & $\zeta$   & sdM, esdM, usdM       \\
                 &     &  \dgr    &   0.1 mag bins
\enddata
 \label{table:subsamples}
\tablecomments{Bins with fewer than 30 stars were not used in this analysis.}
\end{deluxetable}
\end{center}

\begin{center}
\begin{deluxetable}{lr}
\tablewidth{0pt}
\tabletypesize{\small} 
\tablecaption{$(g-r)_{dM}$ Coefficients}
 \tablehead{
 \colhead{Coefficient} &
 \colhead{Value} 
}
 \startdata
$C_0$ & -14.8952 $\pm$ 3.0184 \\ 
$C_1$ & 47.8830  $\pm$    10.1991 \\ 
$C_2$ & -56.3038 $\pm$  13.6807    \\ 
$C_3$ & 34.1232 $\pm$ 9.35810  \\ 
$C_4$ & -11.2926 $\pm$ 3.45636 \\ 
$C_5$ & 1.94407 $\pm$ 0.656087 \\
$C_6$ & -0.136479 $\pm$ 0.0501870 
\enddata
 \label{table:dgr_def}
\tablecomments{Polynomial fit to the $g-r, r-z$ color-color locus of
  SDSS M dwarfs.  Applicable to SDSS stars with $0.8 < r-z < 3.6$.}
\end{deluxetable}
\end{center}

\subsection{Computational Method}
Using the method described in \cite{2011AJ....141...98B} the following
analysis was applied to each subsample.  Ten loops were computed for
each dataset, with 5,000 simplex optimization iterations per loop, for
a total of 50,000 iterations per subsample.
The initial absolute
magnitude estimates for each sample were computed from the $M_r, r-z$ CMR of
\cite{2010AJ....139.2679B}.  After each loop, the absolute
magnitude estimate was updated, and the output of the
previous run was used as input for the next.  Typically,  
convergence was obtained after 5 loops (i.e.,  25,000 iterations).  

\section{Results}\label{sec:results}
There are three major results from our statistical parallax analysis
of subdwarfs: color--absolute magnitude relations, velocity
dispersions and mean motions with respect to the Sun for each
subsample.  We compare our results to previous studies
\citep[i.e.,][]{2011AJ....141...98B} and discuss their relevance to large
surveys of low--mass stars.

\subsection{Absolute Magnitudes}
For stars lacking trigonometric parallax measurements, photometric
colors or spectral types have traditionally been employed to estimate their absolute
magnitudes, resulting in photometric or spectroscopic parallax
relations.  These relations, such as those in
\cite{2002AJ....123.3409H}, \cite{2008ApJ...673..864J} and \cite{2010AJ....139.2679B}
are often based on the colors or spectra of nearby stars with trigonometric
distance estimates.  Deep surveys, such as SDSS, detect low--mass
stars at distances out to a few kpc, which may not share the same age or
chemical abundances as M dwarfs in the solar neighborhood.
\cite{2011AJ....141...98B} demonstrated that age (as traced by H$\alpha$
chromospheric emission) and metallicity (as traced by $\zeta$) can
produce differences of $\sim$ 1 mag in $M_r$ at a given $r-z$ color or
spectral type.  

\begin{figure}[htbp]
  \centering
  \includegraphics[scale=0.4]{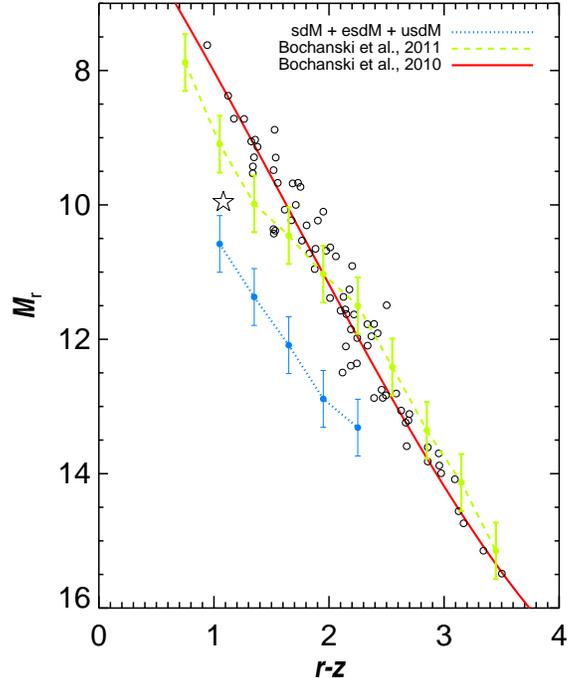}
\caption{The $Mr, r-z$ CMR for nearby stars \citep[open circles and
  red line][]{2010AJ....139.2679B}, 
  the statistical parallax analysis of SDSS M dwarfs \citep[green filled circles and
  dashed line][]{2011AJ....141...98B} and the statistical parallax analysis
  of SDSS subdwarfs (blue filled circles and line).  The subdwarf LP~21-194 is plotted
    as the open star.  Note the
  separation in absolute magnitude between the subdwarfs and disk
  stars at a given $r-z$ color.}
  \label{fig:all}
\end{figure}

In Figure \ref{fig:all}, we compare the statistical parallax results
from the current analysis and \cite[][green line]{2011AJ....141...98B} to the nearby star
photometry.  The $M_r, r-z$
CMR measured by \cite{2010AJ....139.2679B} for nearby,
solar--metallicity stars is shown with the red line.
Included in Figure \ref{fig:all} is the subdwarf LP~21-194, which was
observed as part of the nearby star sample
\citep{bochanskithesis}. The subdwarf sample from the current study
(blue line) shows a systematic shift
compared to the main--sequence M dwarf statistical parallax results and the
nearby stars.   As first noted by \cite{1959MNRAS.119..278S}, low--metallicity
subdwarfs lie below the main sequence of stars, with a fainter
absolute magnitude at the same color.   The subdwarf sample does not
span a comparable color range as the \cite{2011AJ....141...98B} results,
since there are fewer stars to populate each color bin.

To further quantify the dependence of absolute magnitude on metallicity,
we calculated statistical parallaxes in the $r,i$ and $z$ bands. Both
$\zeta$ and {\dgr} were used to trace metallicity, and the results are
discussed below.  

\subsubsection{Absolute Magnitude Variations and $\zeta$}
The left panels of Figures \ref{fig:mr_zeta_dgr},
\ref{fig:mi_zeta_dgr}, and \ref{fig:mz_zeta_dgr} show the derived CMRs
for each metallicity subclass (sdM, esdM and usdM).  The three
classes are compared to nearby M dwarfs and the best--fit nearby star CMR from
\cite{2010AJ....139.2679B}.  The CMR for each subclass is roughly
parallel to the main sequence, with an offset of $\sim$ 1 mag in
absolute magnitude between adjacent subclasses at the same $r-z$ color.  The largest separation
from the main sequence occurs for the most metal--poor subclass,
usdM.  For example, at $r-z \sim 1.35$, there are $\sim$ 3 mags
difference between the main sequence and the usdM CMR.  At a fiducial
apparent magnitude of $r \sim 18$, this can result in overestimating the
distance by a factor of $\sim 5\times$.  Thus, identifying the
metallicity subclass is crucial for determining precise stellar distance
estimates.  Despite the large spectroscopic campaigns planned for the
near future such as GAIA and 4MOST \citep{2001A&A...369..339P, 2012arXiv1206.6885D}, the
vast majority of subdwarfs will only have $ugriz$ photometry, courtesy of
LSST. 

\subsubsection{Absolute Magnitude Variations and {\dgr}}

A photometric proxy of metallicity for M dwarfs is complicated by the
lack of a comprehensive \textit{spectroscopic} calibration.  However,
the outlook is promising, as recent investigations have resulted in spectroscopic metallicities of
low--mass stars with $\sim 0.1$ dex precision, over limited ranges.
These studies have focused mainly on the infrared \citep{2012ApJ...748...93R}, but some optical metallicity calibrations \citep{2012arXiv1211.4630M} have
been developed.  The {\dgr} metric was designed to use the 
$ugriz$ photometric system.  We eschewed the use of $u$, as the SED of
M dwarfs is peaked near the near--infrared, with very little $u$--band
flux.  As noted in \cite{2004AJ....128..426W}, the $g-r$ colors of subdwarfs are
redder at a given spectral type or $r-i$ color, which prompted the
construction of the {\dgr} index.  Both $\zeta$ and {\dgr} are crude
metallicity indicators, but they do trace the same behavior, as shown
in  Figure \ref{fig:dgr_zeta}.  There is a correlation between the two values,
albeit with large scatter, with metal--poor values of $\zeta$ corresponding
to larger {\dgr} values.   

\begin{figure*}[htbp]
  \centering
 \begin{center}$
\begin{array}{cc}  
 \includegraphics[scale=0.4]{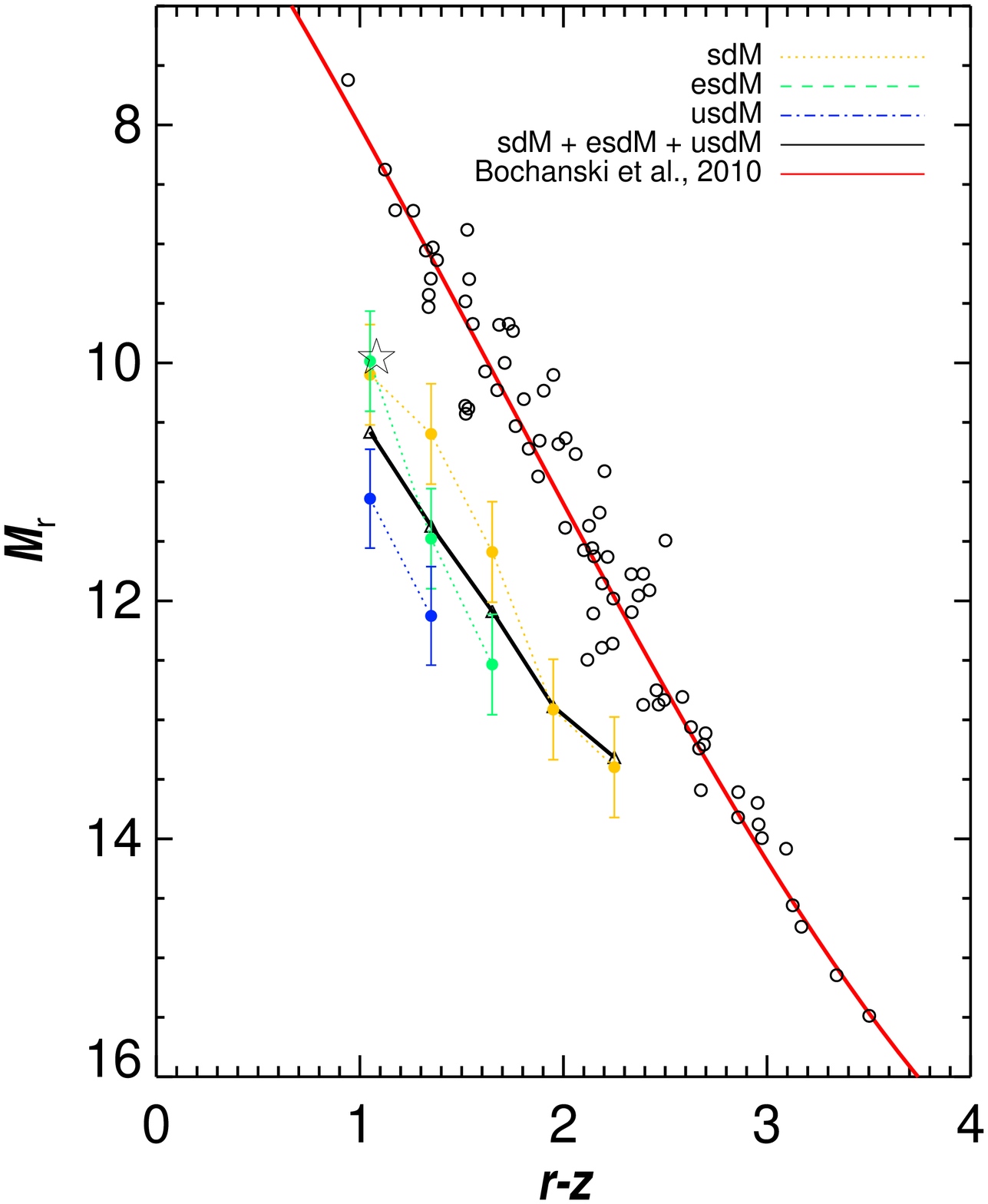}&
  \includegraphics[scale=0.4]{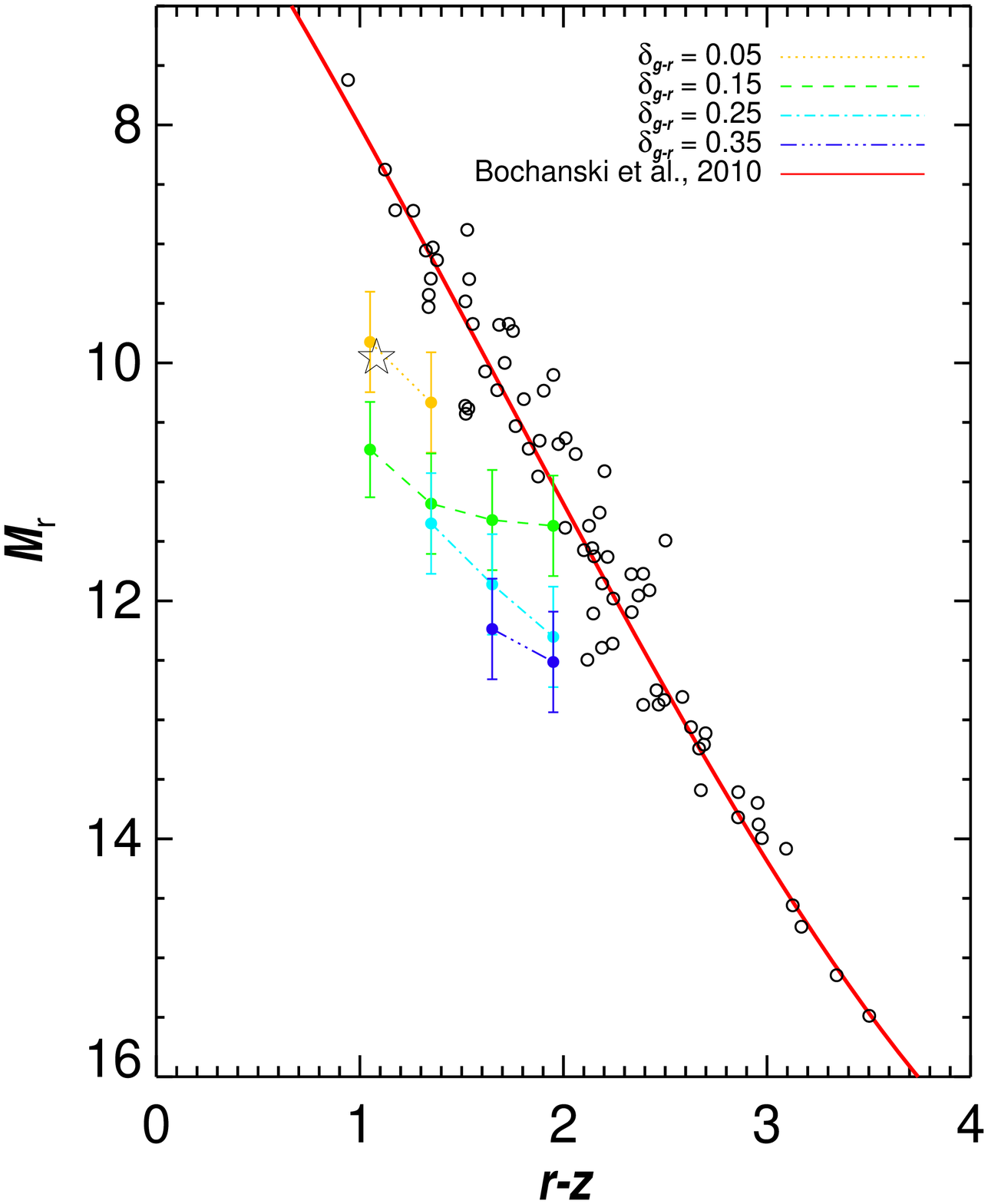} 
\end{array}$
\end{center}  
\caption{Left Panel: $M_r$ vs. $r-z$ for various $\zeta$ cuts.   The open circles are nearby
    stars with accurate trigonometric parallaxes from
    \cite{bochanskithesis} with the best--fit relation of
    \cite{2010AJ....139.2679B} shown in red.  The subdwarf LP~21-194 is plotted
    as the open star.  The solid black line denotes
    the entire subdwarf sample, with various $\zeta$ cuts described by the
    legend.  The spectral metallicity classes are well behaved, with
    clear separation for most color bins.   
  Right Panel: $M_r$ vs. $r-z$ as a function of {\dgr}, with same symbols and lines as in
the spectral type panel.  The separation between adjacent {\dgr} CMRs
is smaller, indicating sampling of metallicity--absolute magnitude space.}
  \label{fig:mr_zeta_dgr}
\end{figure*}

\begin{figure*}[htbp]
  \centering
 \begin{center}$
\begin{array}{cc}  
  \includegraphics[scale=0.4]{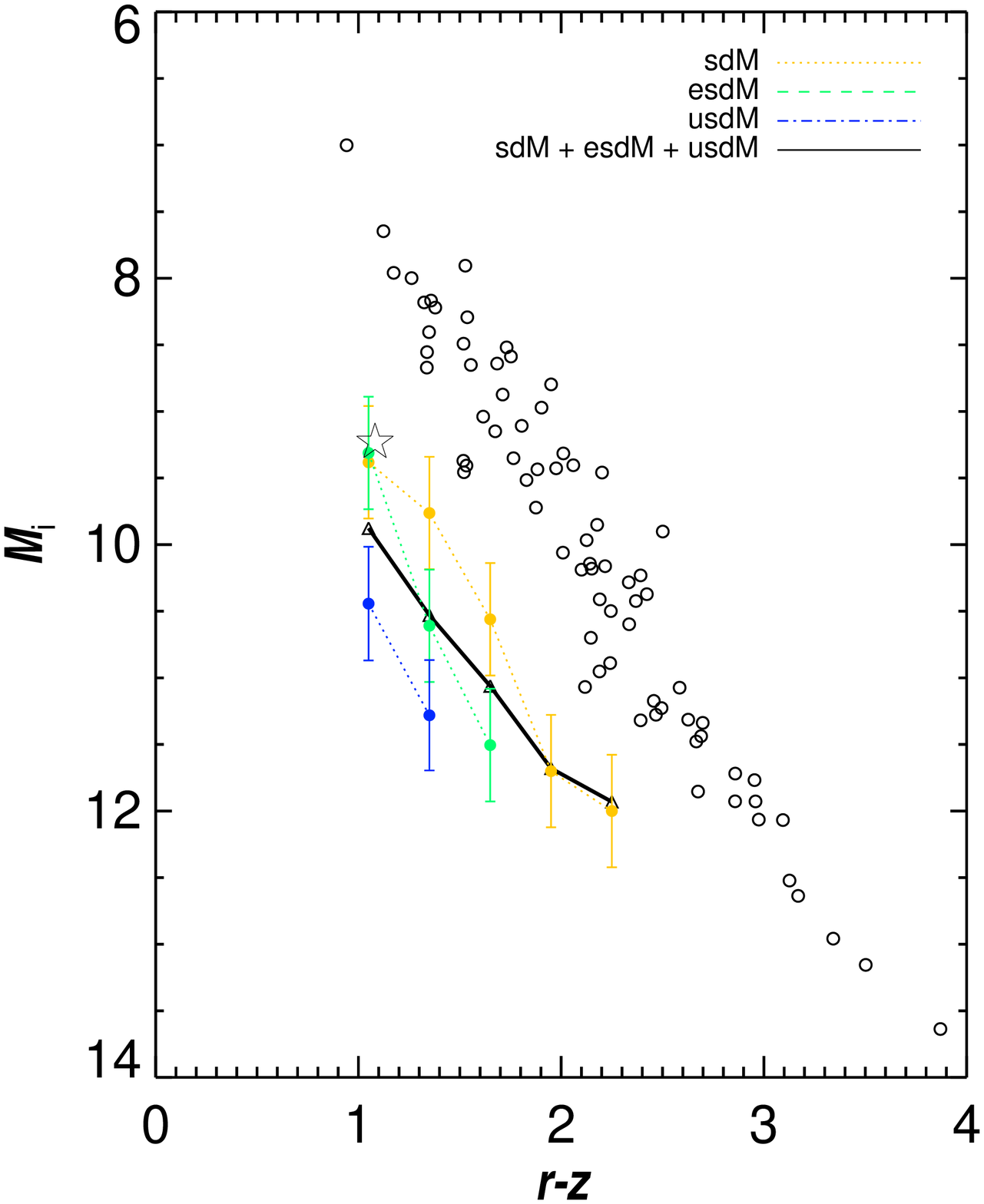}&
  \includegraphics[scale=0.4]{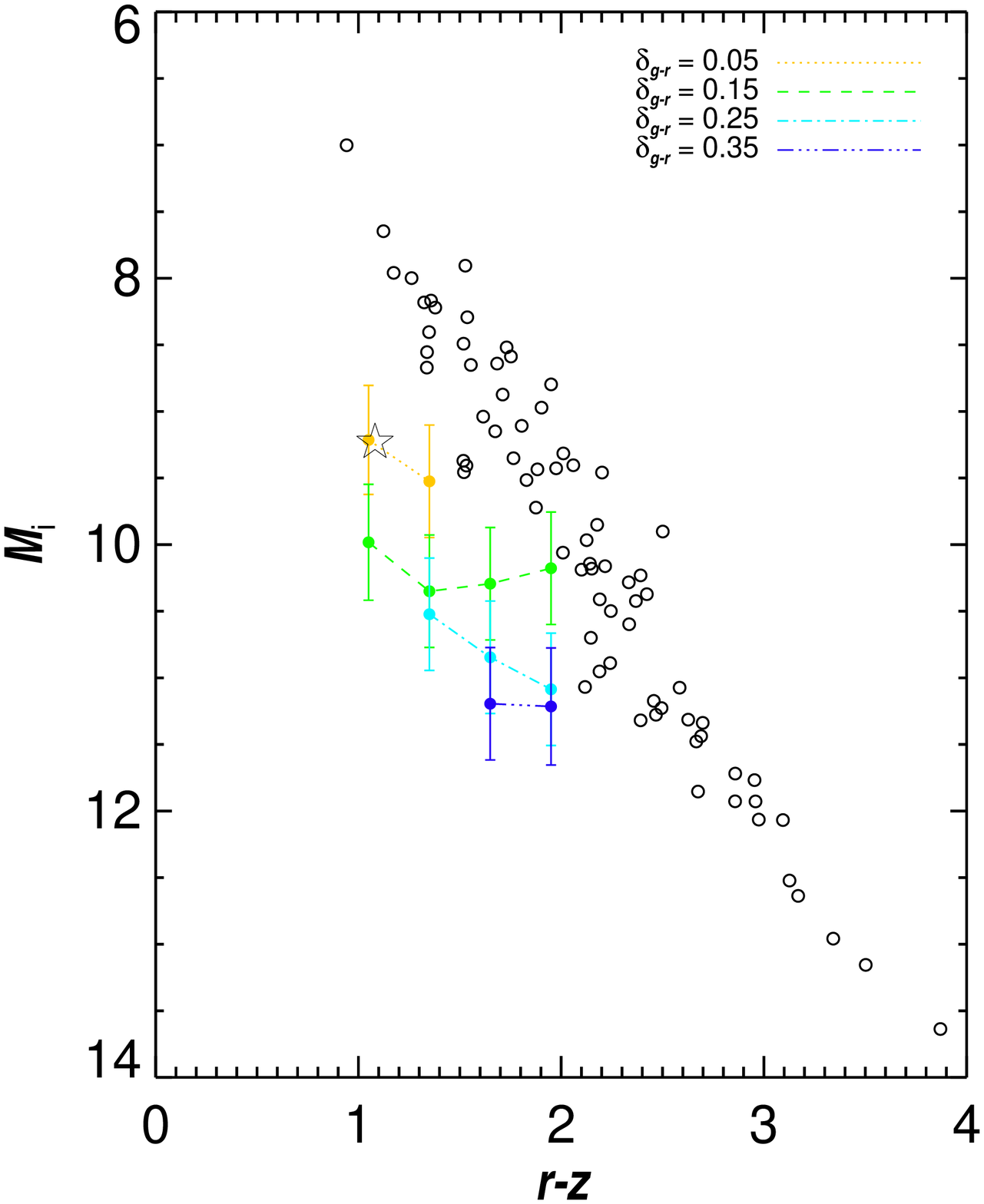} 
\end{array}$
\end{center}  
\caption{Left Panel: $M_i$ vs. $r-z$ for various $\zeta$ cuts.   The
  lines and symbols follow Figure \ref{fig:mr_zeta_dgr}.
  Right Panel: $M_i$ vs. $r-z$ as a function of {\dgr}, with same symbols and lines as in
the spectral type panel.}
  \label{fig:mi_zeta_dgr}
\end{figure*}

\begin{figure*}[htbp]
  \centering
 \begin{center}$
\begin{array}{cc}  
  \includegraphics[scale=0.4]{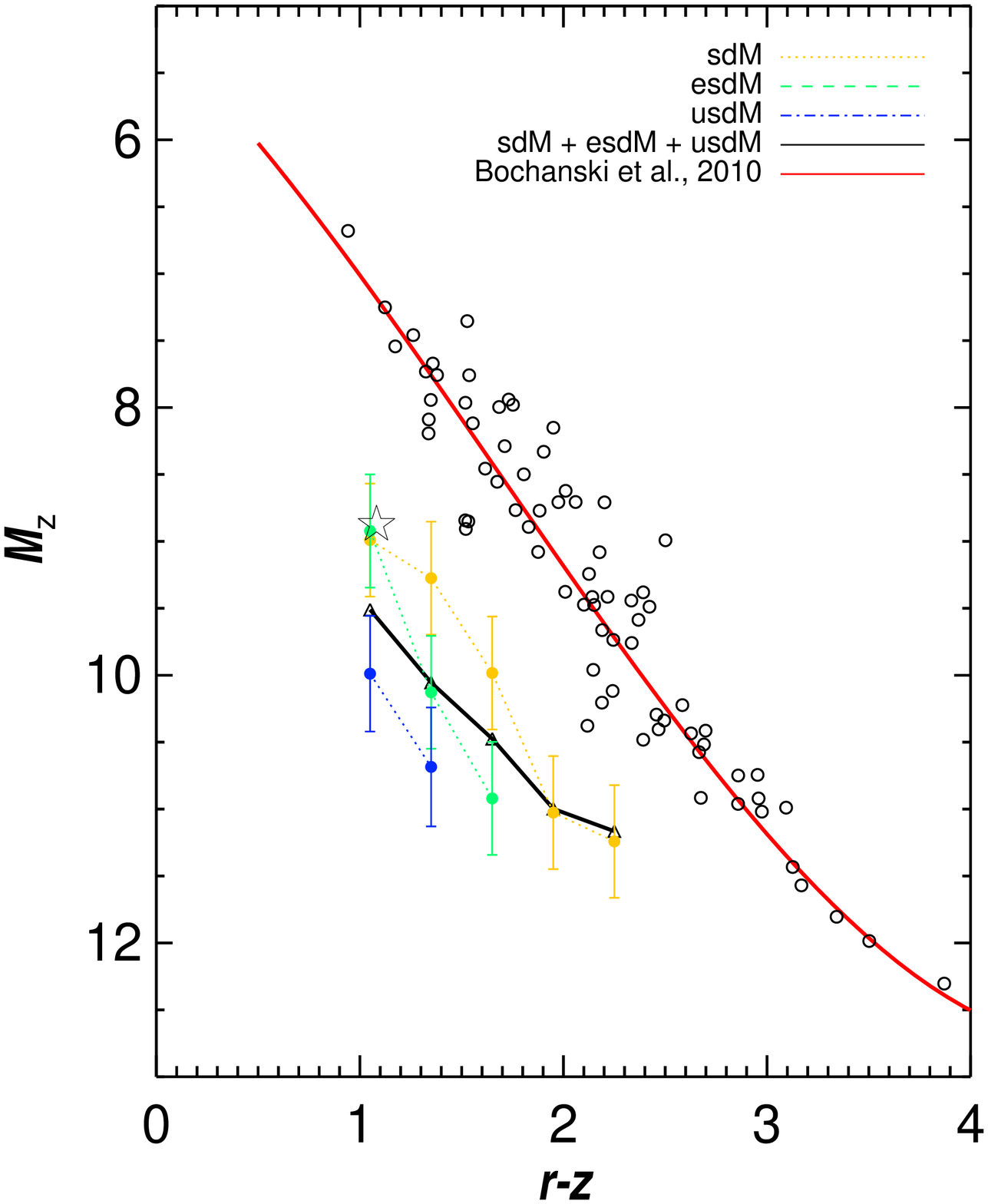}&
  \includegraphics[scale=0.4]{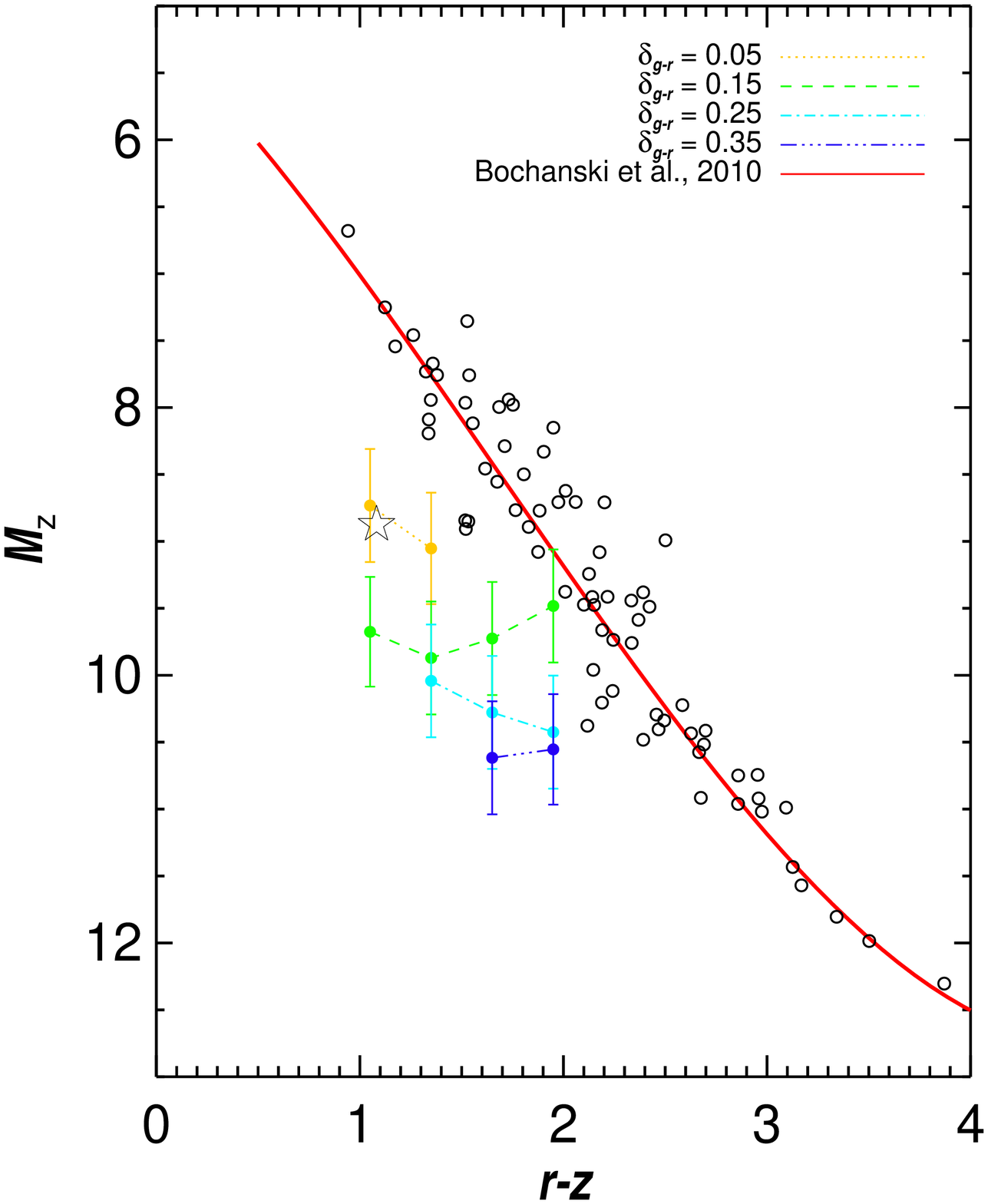} 
\end{array}$
\end{center}  
\caption{Left Panel: $M_z$ vs. $r-z$ for various $\zeta$ cuts.   The
  lines and symbols follow Figure \ref{fig:mr_zeta_dgr}.
  Right Panel: $M_z$ vs. $r-z$ as a function of {\dgr}, with same symbols and lines as in
the spectral type panel.}
  \label{fig:mz_zeta_dgr}
\end{figure*}

\begin{figure}[htbp]
  \centering
  \includegraphics[scale=0.4]{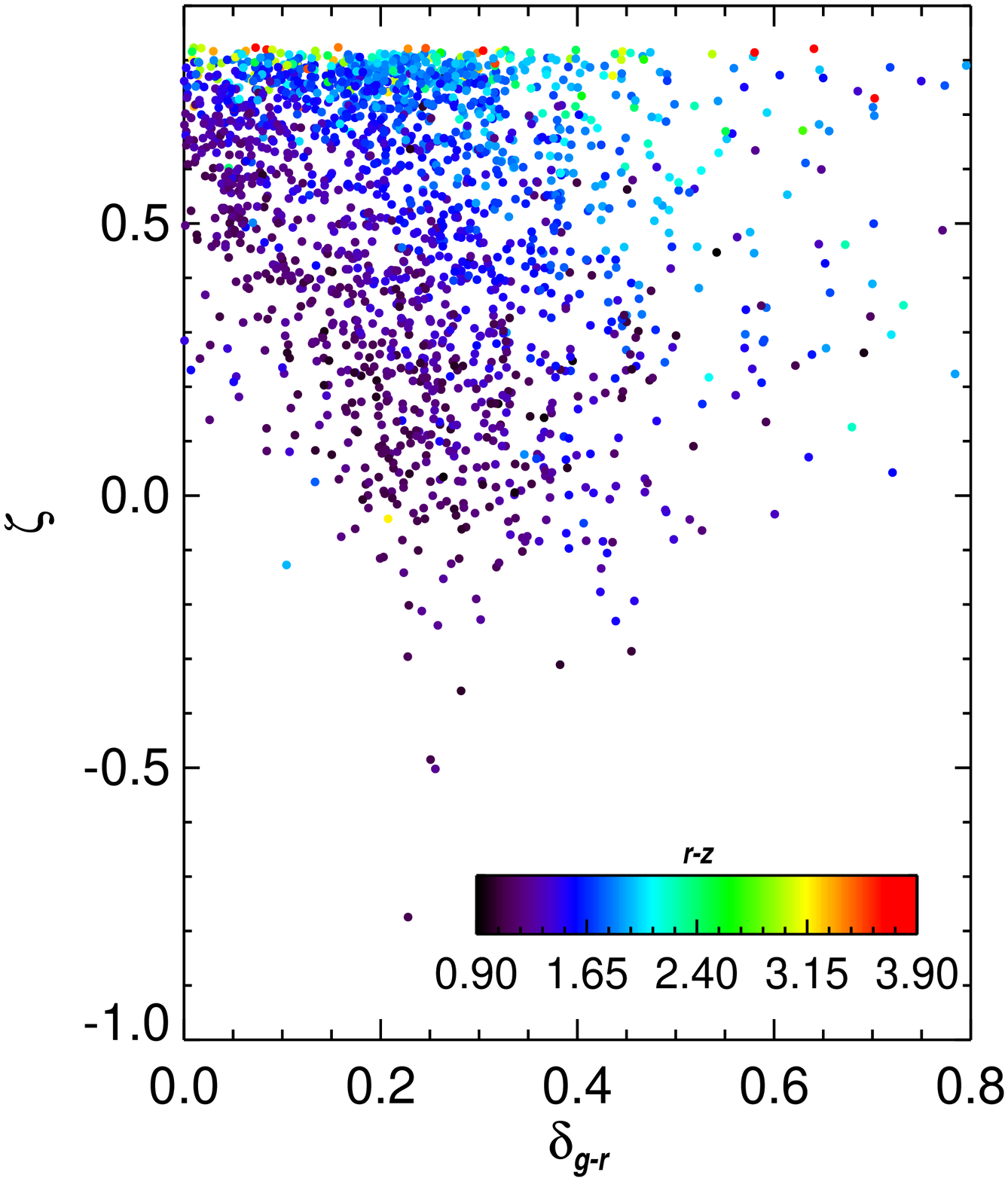}
\caption{$\zeta$ vs. {\dgr} for the subdwarf sample.  Note that {\dgr}
roughly traces $\zeta$, with small values of $\zeta$ corresponding to
larger (redder) values of {\dgr}.  Each point is color--coded by the
star's $r-z$ color according to the color bar.}
  \label{fig:dgr_zeta}
\end{figure}

The right panels of Figures \ref{fig:mr_zeta_dgr},
\ref{fig:mi_zeta_dgr}, and \ref{fig:mz_zeta_dgr} show the CMRs derived
as a function of {\dgr}.  The {\dgr} CMRs share similar trends when
compared to the CMRs derived for different metallicity subclasses.
Stars with larger {\dgr} values are intrinsically fainter, as expected
for more metal--poor stars. The separation between {\dgr} loci is
smaller, suggesting a finer sampling of the metallicity--absolute
magnitude phase space.

We derived photometric parallax relations by fitting the $M_{riz},
r-z$, {\dgr} plane, with the following functional form:
\begin{eqnarray}
M_{r,i,z} = a_0  + a_1 \times (r-z) + a_2 \times (r-z)^2  
\nonumber\\+  a_3 \times \delta_{(g-r)} + a_4 \times (r-z) \times \delta_{(g-r)},
\label{eqn:phots}
\end{eqnarray}
\setcounter{equation}{3}
\noindent where $a_0 ... a_4$ are given in Table
\ref{table:abs_mags}.  We selected simple functions:  low--order
polynomials and cross--terms to derive the relations, while still
producing good fits for each relation.  The standard deviation between
the data and each fit was $\sim 0.1$ mag.  Combined with the
uncertainty of each point $\sim 0.4$ mag, the total uncertainty is
0.41 mag.  The functions, standard deviations and applicable color
ranges are listed in Table \ref{table:abs_mags}.  These parallax
relations are most appropriate for SDSS observations, but should be
applicable to any calibrated $ugriz$ photometry.

\begin{turnpage}
\begin{center}
\begin{deluxetable*}{ccccccll}
\tablewidth{0pt}
\tabletypesize{\scriptsize}
\tablecaption{Absolute Magnitudes of SDSS M Dwarfs}
\tablehead{
\colhead{Absolute Mag.}&
\multicolumn{5}{c}{Coefficients\tablenotemark{a}} &
\colhead{$\sigma_{\rm Tot}$\tablenotemark{b}}&
\colhead{Color Range}
}

\startdata
          &     $a_0$    &    $a_1$    &    $a_2$    &    $a_3$    &         $a_4$    &  & \\ 
\hline
$M_r$ &  7.9547 $\pm$  1.578&    1.8102 $\pm$ 2.294&     -0.17347$\pm$0.814&  7.7038 $\pm$ 4.511&      -1.4170 $\pm$ 2.771 &  0.41 &  $ 1.0 < r-z < 2.0; 0.0 <$ {\dgr} $< 0.5$ \\
$M_i$ &  7.3478 $\pm$ 1.503&      2.0068 $\pm$ 2.189&    -0.45371$\pm$ 0.776&  7.0445   $\pm$ 4.249 &   -1.2128  $\pm$2.612&  0.41 &  $ 1.0 < r-z < 2.0; 0.0 <$ {\dgr} $< 0.5$\\
$M_z$ &  7.3798 $\pm$ 1.6723&      1.4977 $\pm$ 2.430&     -0.33956 $\pm$  0.862&  8.4248 $\pm$4.738 &     -1.9804 $\pm$2.908&  0.42 &  $ 1.0 < r-z < 2.0; 0.0 <$ {\dgr} $< 0.5$
\enddata
 \label{table:abs_mags}
\tablenotetext{a}{The functional form for the photometric parallax is
  given in Equation \ref{eqn:phots}.}
\tablenotetext{b}{The uncertainty is given by $\sigma_{\rm Tot} = \sqrt{\sigma_{\rm fit}^2 +
    \sigma_M^2}$, where $\sigma_M \sim 0.4$ and $\sigma_{\rm fit} \sim
  0.1$.}
\tablecomments{The typical uncertainty for the reported absolute
  magnitudes is $\sigma_{M} = 0.42$, which is directly computed from
  Equation \ref{eqn:sigk}.}
\end{deluxetable*}
\end{center}
\end{turnpage}

\subsection{Kinematics of M Subdwarfs}\label{sec:kinematics}
In addition to estimating the absolute magnitude of a homogeneous set
of stars, the statistical parallax method also measures the velocity
ellipsoid (dispersions and orientation) and the mean motion of the
population.  The mean motions are reported as the Sun's peculiar
velocity with respect to the statistical parallax targets.  We
compared the mean motions of the subdwarf sample to disk M dwarfs \citep{2011AJ....141...98B} as
a function of $r-z$ color.

\subsubsection{Solar Peculiar Motion}
The reflex motion of the Sun is reported by the statistical parallax
analysis, measured with respect to the mean velocity of the
subsample.  Subsamples with different mean velocities will reflect
different values for the Sun's peculiar motion.  We
employ a right--handed coordinate system with $U$ increasing towards the Galactic
center, $V$ increasing in the direction of solar motion, and $W$
increasing vertically upward (as in \citealp{Dehnen98}).  In this
coordinate system the angular momentum vector of the solar
orbital motion points towards the
South Galactic Pole.  

\begin{figure*}[htbp]
  \centering
 \begin{center}$
\begin{array}{cc}  
  \includegraphics[scale=0.4]{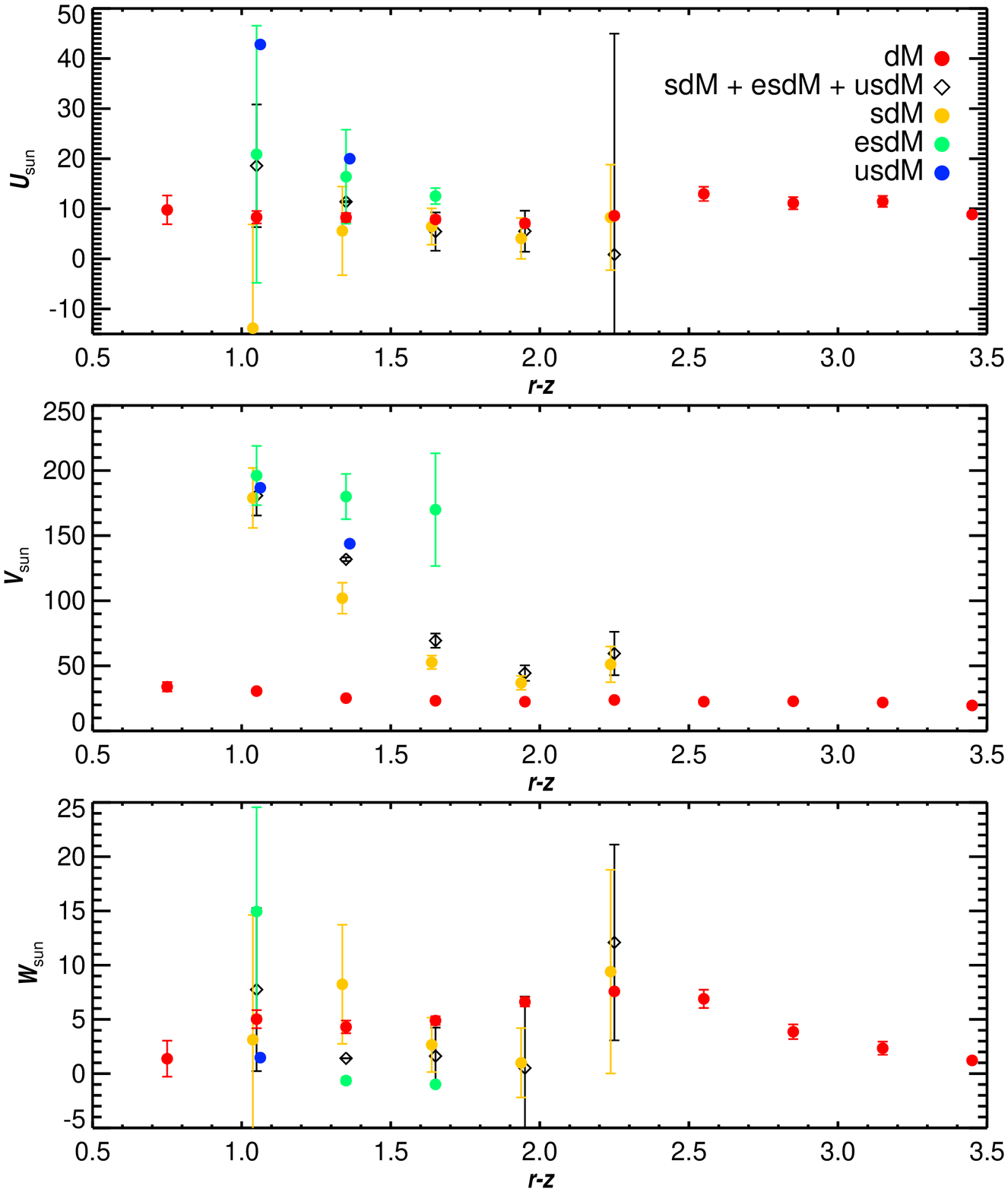}&
  \includegraphics[scale=0.4]{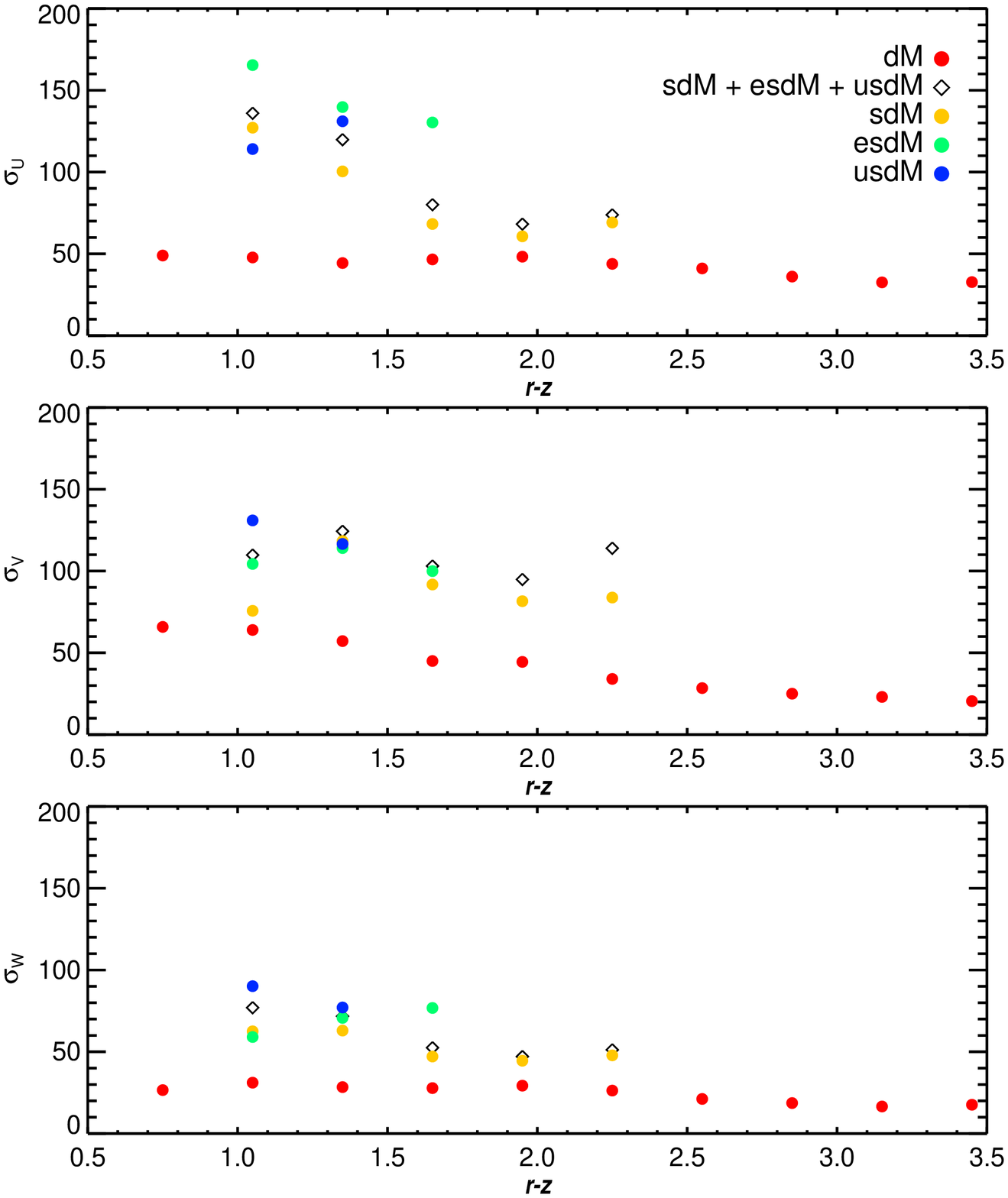} 
\end{array}$
\end{center}  
\caption{Left Panels: The mean velocities of each metallicity subclass as
  a function of $r-z$ color.  The total sample is marked by the open
  diamonds, and disk M dwarfs are shown with the red filled circles.
  Each metallicity class is described in the legend.  While the disk M
  dwarfs have a smaller mean $V$ velocity (middle panel), the
 bluest subdwarf bin has $V \sim 200$ {\kms}, indicative of a
 non--rotating population.
  Right Panels: Velocity dispersions along the $UVW$ axes for each
  metallicity subclass as a function of $r-z$ color.  Symbols are the
  same as the left panels.  Note the larger velocity dispersions at
  all color bins for the subdwarf population, indicative of an older,
  dynamically heated population.}
  \label{fig:kinematics}
\end{figure*}

In the left panel of Figure \ref{fig:kinematics} we compare the mean
motions of the metallicity subclasses as a function of $r-z$ color.
The mean solar motion measured with respect to disk M dwarfs is also
included \citep{2011AJ....141...98B}.   For most color bins, the motion of the
entire subdwarf sample (sdM + esdM + usdM) closely mimics the motions
of the sdM population, since most of the stars in the sample are found
in this subclass.  The $U$ direction shows a large dispersion among
the subclasses (i.e., $r-z = 1.05$), but with general agreement
between the subdwarf and main--sequence populations, indicating
similar radial motions through the Galaxy.  The $W$ direction exhibits
large dispersions within individual color bins as well, but the entire
subdwarf population shows agreement with the disk M dwarfs at the
$\sim$ 5 {\kms} level.   

The most
significant contrast between the disk M dwarfs and subdwarfs is
manifested in the $V$ direction, where the Sun's reflex motion is near
200 {\kms} for the bluest subdwarf subsamples.  Past investigations of
subdwarf motions 
\citep[i.e., ][]{1997AJ....113..806G} have measured $\langle~V~\rangle
\sim 200$ {\kms}.   This is very similar to
the Sun's circular velocity around the Galactic center \citep[220{\kms}; ][]{1986MNRAS.221.1023K}.  Thus, the
subdwarfs comprise a population with \textit{very little} ordered
rotation, indicative of halo orbits.  These orbits are characterized
by high eccentricity and inclination.   Orbits for ultra--cool subdwarfs
(spectral types $>$ M7) have
been a recent topic of investigation \citep{2009ApJ...696..986C, 2008ApJ...672.1159B,2009ApJ...697..148B},
since kinematics may be useful for constraining ages of brown dwarfs,
or identifying new tidal debris streams \citep{2006ApJ...642L.137B}.  At redder
colors (and closer distances), the average $V$ velocity of the
subdwarf population falls to $\sim$ 60 {\kms}.  This velocity is
significantly larger than nearby M dwarfs \citep[24 $\pm 3$ {\kms}; ][]{2011AJ....141...98B}, but smaller
than the bluest subdwarfs.  These stars are probably members of the thick
disk, while the bluer subdwarfs are primarily halo constituents.  The
mean velocities of each subsample can be found in Table \ref{table:vels}.

\subsubsection{Velocity Dispersions}
Subdwarfs comprise a kinematically hotter population than disk M
dwarfs, as shown by their larger velocity dispersions.  In the right
panels of Figure \ref{fig:kinematics}, we compare the subdwarf
metallicity classes to the M dwarf sample \citep{2011AJ....141...98B} as a
function of $r-z$ color.   Some immediate trends are obvious.  First,
the total subdwarf sample (open diamonds) has larger velocity
dispersions at all color bins, when compared to the M dwarf sample
(red filled circles).  
While velocity dispersions within the disk are thought to
increase roughly as the square--root of age
\citep[e.g.,][]{2001ASPC..228..235F,2002MNRAS.337..731H} due to
disk heating, the halo
dispersion is governed by the Galactic potential.
This implies that velocity dispersions do not closely follow the
relative age differences.   However, both observational \citep{2006ApJ...638..446M} and
theoretical investigations \citep{2009ApJ...702.1058Z, 2012arXiv1206.0740B} demonstrate that halo
subdwarfs comprise an  
 older, metal--poor population than nearby disk stars.  The bluer $r-z$ bins contain subdwarfs with
dispersions of $(\sigma_U, \sigma_V, \sigma_W) \sim (110, 100, 70)$
{\kms}.   The typical
uncertainty in each dispersion measurement is $\sim 5-10$ {\kms}.  These dispersions suggest the stars are drawn from the inner
halo \citep{2010ApJ...712..692C}.    Moreover, the radial ($U$) and
vertical ($W$) dispersions increase with decreasing metallicity for
most bins, following trends seem in other studies \citep{2010ApJ...712..692C,2010ApJ...716....1B}.   Similar to the M dwarfs, the smallest velocity
dispersions are seen in the reddest three $r-z$ bins in the $W$
direction, indicating that subdwarfs with these colors may possess
orbits with some ordered rotation, and suggesting that these stars
are drawn from the thick disk.  In Table \ref{table:disps}, we
list the velocity dispersions for each spectral subclass as a function
of $r-z$ color, as well as the disk M dwarfs from \cite{2011AJ....141...98B}.

\subsubsection{Probability Plots}
\begin{figure}[htbp]
  \centering
  \includegraphics[scale=0.4]{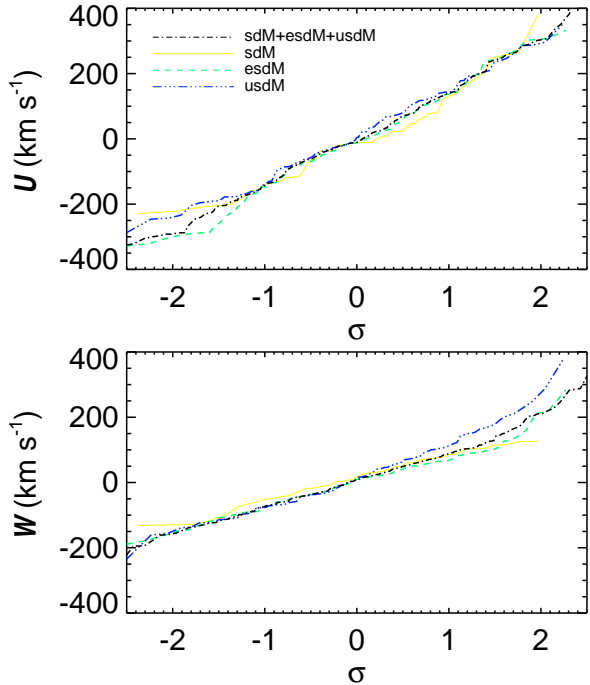}
\caption{Probability plots for the $U$ and $W$ velocity distributions
  of the $r-z = 1.05$ bin.  Each line is described in the legend.  The
  various metallicity subclasses are well described by a single
  component, validating our assumption of kinematic homogeneity within
  each subsample.  The $V$ velocity probability plot was omitted since
  the distribution is non-Gaussian due to asymmetric drift, an effect
  which is included in the statistical parallax analysis.
}
  \label{fig:prob_plots}
\end{figure}

The statistical parallax method assumes that each subsample is
composed of a kinematically homogenous population which can be
described by a velocity ellipsoid.  Some metal--poor stellar samples
have contained stars from the thick disk and halo, which possess
different underlying kinematics \citep{2007AJ....134.2418B,2010ApJ...712..692C}.  If a given subsample was comprised
of similar numbers of thick disk and halo subdwarfs, a simple one
component fit would not be valid.  We constructed probability plots
\citep{1980AJ.....85.1390L,2002AJ....124.2721R} to test the validity of this assumption.  Space motions
were calculated from the proper motions and radial velocities from
SDSS, along with distances estimated from the $M_r$ relation in
Equation \ref{eqn:phots}.  An illustrative
example of the probability plots are shown in Figure
\ref{fig:prob_plots}. The $U$ and $W$ probability plots for
the $r-z = 1.05$ bin for the various metallicity subclasses are plotted.
Probability plots are related to cumulative distribution functions,
with the abscissae calculated in units of standard deviation.  Hence,
a Gaussian distribution would be plotted as a line, with a slope equal
to the dispersion. Populations consisting of multiple Gaussians would be manifested as a segmented line (see Figure
3 of \citealp{2007AJ....134.2418B} for an example).  The probability plots
show that the underlying distributions are well described by a single
component, validating our assumption.  

\section{Conclusions}\label{sec:conclusions}
We have described the statistical parallax analysis of low--mass
subdwarfs contained in the SDSS catalog \citep{antonia}.  We divided
our sample by color, metallicity subclass, and photometric properties
to further explore the kinematics and intrinsic brightnesses of these
metal--poor stars.  We developed a photometric proxy of metallicity
for low--mass subdwarfs, derived photometric parallax relations, and 
explored the mean motions and dispersions
with respect to their metal--rich counterparts.

Metallicity can influence the
color--magnitude relations of low--mass stars.  While this has been
directly measured for more massive stars in the $ugriz$ photometric
system \citep{2008ApJS..179..326A, Ivezic08}, there has not been a detailed study for
low--mass stars.  We expanded on the
initial analysis of \cite{2011AJ....141...98B} to include subdwarfs with
lower metallicity, as measured by the proxy $\zeta$.   We demonstrated
that different metallicity subclasses (sdM, esdM and usdM) cleanly separated in various color--magnitude
diagrams.  The one nearby subdwarf with a precise
trigonometric parallax and $ugriz$ photometry agrees with our derived
relations.

While deriving color--absolute magnitude relations for various
spectral subclasses does demonstrate the influence of metallicity on
the intrinsic brightness of stars, it requires that spectroscopic
observations are available.  The largest current and future surveys, such as PanStarrs
and LSST, will be purely photometric, limiting the application of the
spectral type relations.  We ameliorated this situation by deriving
{\dgr}, a photometric proxy for metallicity based on a star's $g-r$
and $r-z$ colors.  Low--mass subdwarfs have redder $g-r$ colors than
their solar--metallicity counterparts \citep{2004AJ....128..426W} and we quantified
this effect by measuring the $g-r, r-z$ color--color locus for main
sequence M dwarfs.   The photometric metallicity parameter {\dgr},
reasonably traces the spectroscopically determined $\zeta$
values for low--mass subdwarfs, with some scatter, and complements
earlier investigations by \cite{2011AJ....141...97W} for solar metallicity M dwarfs.   These relations have
potential to be a powerful tool in photometrically identifying
low--mass subdwarfs.

The kinematic properties of subdwarfs were also explored in this
study.  The entire sample was compared to disk M dwarfs, and some
clear differences were seen.  First, the mean $V$ velocity of bluer
subdwarfs is near 200 {\kms}, suggesting this population is largely
non--rotating.  These objects possess halo--type orbits, with
high eccentricities and inclinations.  The velocity dispersions
of the bluest (in $r-z$) subdwarfs was similar to other measurements
of the Milky Way's inner halo.  The redder $r-z$ bins contained stars
with kinematics similar to the thick disk.

The Milky Way's halo has yet to be mapped by its 
most numerous inhabitant, low--mass subdwarfs.  While existing
surveys, such as SDSS and 2MASS have offered a small glimpse into the
low--mass components of the halo, this phase space is largely
unexplored.  The relations derived in this study will enable easier
identification and characterization of these objects in the next
generation of astronomical surveys, such as LSST and GAIA.  These
surveys will probe the halo \textit{in situ} and offer the most detailed
picture of its kinematic and structural composition.

\acknowledgements
We thank Beth Willman, Steve Boughn and Pat Boeshaar for helpful discussions. 
AAW acknowledges the financial support of NSF grant AST 11-09273.
We also gratefully acknowledge the support of NSF grants AST 02-05875
and AST 06-07644 and NASA ADP grant NAG5-13111.   

Funding for the SDSS and SDSS-II has been provided by the Alfred P. Sloan Foundation, the Participating Institutions, the National Science Foundation, the U.S. Department of Energy, the National Aeronautics and Space Administration, the Japanese Monbukagakusho, the Max Planck Society, and the Higher Education Funding Council for England. The SDSS Web Site is http://www.sdss.org/.

The SDSS is managed by the Astrophysical Research Consortium for the Participating Institutions. The Participating Institutions are the American Museum of Natural History, Astrophysical Institute Potsdam, University of Basel, University of Cambridge, Case Western Reserve University, University of Chicago, Drexel University, Fermilab, the Institute for Advanced Study, the Japan Participation Group, Johns Hopkins University, the Joint Institute for Nuclear Astrophysics, the Kavli Institute for Particle Astrophysics and Cosmology, the Korean Scientist Group, the Chinese Academy of Sciences (LAMOST), Los Alamos National Laboratory, the Max-Planck-Institute for Astronomy (MPIA), the Max-Planck-Institute for Astrophysics (MPA), New Mexico State University, Ohio State University, University of Pittsburgh, University of Portsmouth, Princeton University, the United States Naval Observatory, and the University of Washington.

\begin{turnpage}
\begin{center}
 \tabletypesize{\scriptsize}
\begin{deluxetable}{r|rrrrr|rrrrr|rrrrr}
\tablewidth{0pt}
 \tablecaption{Mean Motions of M subdwarfs}
\tablehead{
 \colhead{$r-z$} &
 \multicolumn{5}{c}{$U_{\odot}$ (km s$^{-1}$)}&
\multicolumn{5}{c}{$V_{\odot}$ (km s$^{-1}$)}&
\multicolumn{5}{c}{$W_{\odot}$ (km s$^{-1}$)}
}
 \startdata
	& 	sdM	& 	esdM 	& 	usdM  &	Total	&        Disk 	& 	sdM	& 	esdM 	& 	usdM  &	Total     & 	Disk  &	sdM	& 	esdM 	& 	usdM  &	Total	& 	Disk \\
\hline
1.05 &  -14 $\pm$ 21 & 21 $\pm$ 26 & 43 $\pm$ 1  & 19  $\pm$ 12&  8& 179  $\pm$ 23 & 196 $\pm$ 23 & 187  $\pm$ 1 & 180 $\pm$ 15  & 30 & 3 $\pm$ 11 & 15 $\pm$ 10  & 2 $\pm$ 1 & 8  $\pm$ 8& 5 \\
1.35 &  6 $\pm$ 9 &  16 $\pm$ 9 & 20 $\pm$ 1 & 11  $\pm$ 1& 8   & 102  $\pm$ 12 & 180 $\pm$ 17 & 144 $\pm$ 1  &  132 $\pm$ 1 & 25 & 8 $\pm$ 5 & -1 $\pm$ 1 &-18 $\pm$ 1 & 1 $\pm$ 1  & 4 \\
1.65  &  6 $\pm$ 4 & 13 $\pm$ 2 &  \nodata  &  5 $\pm$ 3 & 8   &53  $\pm$ 5  &170 $\pm$ 43  &\nodata  & 69 $\pm$ 5 & 23 &3 $\pm$ 3  &-1 $\pm$ 1  &\nodata  & 2 $\pm$ 3 & 5 \\
1.95  &   4 $\pm$ 4 &  \nodata  &\nodata  & 5 $\pm$ 4 & 7  &37  $\pm$5  &\nodata  &\nodata  & 44  $\pm$ 6  & 22 &1 $\pm$ 3  &\nodata  &\nodata  & 1 $\pm$ 7 & 7 \\
2.25  &  8 $\pm$ 10 &    \nodata  &\nodata  & 0 $\pm$ 44 & 8 &51$\pm$ 14  &\nodata  &\nodata  & 59 $\pm$ 16 & 23 &9 $\pm$ 9  &\nodata&\nodata  & 12 $\pm$ 9 & 7
\enddata
\label{table:vels}
\tablecomments{We employ a right--handed coordinate system with $U$ increasing towards the Galactic
center, $V$ increasing in the direction of solar motion, and $W$
increasing vertically upward (as in \citealp{Dehnen98}). }
\end{deluxetable}
\end{center}
\end{turnpage}

\begin{center}
\begin{deluxetable}{r|rrrrr|rrrrr|rrrrr}
\tablewidth{0pt}
 \tablecaption{Velocity Dispersions of M subdwarfs}
 \tabletypesize{\scriptsize}

 \tablehead{
 \colhead{$r-z$} &
 \multicolumn{5}{c}{$\sigma_U$ (km s$^{-1}$)}&
\multicolumn{5}{c}{$\sigma_V$ (km s$^{-1}$)}&
\multicolumn{5}{c}{$\sigma_W$ (km s$^{-1}$)}
}
 \startdata
	& 	sdM	& 	esdM 	& 	usdM  &	Total	&        Disk 	& 	sdM	& 	esdM 	& 	usdM  &	Total     & 	Disk  &	sdM	& 	esdM 	& 	usdM  &	Total	& 	Disk \\
\hline
1.05 &  127 & 165 &  114 & 136 &  48    &76  &104  &131  &110  &64  &63  &59  &90  &77  &31  \\
1.35 &  100 &  140 & 131 &  120 & 44    &118  &114  &117  &124  &57  &63  &71  &77  &72  &29  \\
1.65  &  68 & 130 &  \nodata  & 80  &47    &92  &100  &\nodata  &103  &45  &47  &77  &\nodata  &53  &28  \\
1.95  &   61 &  \nodata  &\nodata  & 68 &  48  &82  &\nodata  &\nodata  &95  &44  &45  &\nodata  &\nodata  &47  &29  \\
2.25  &  69 &    \nodata  &\nodata  &  74 & 44  &84  &\nodata  &\nodata  &114  &34  &48  &\nodata  &\nodata  &51  &26  
\enddata
 \label{table:disps}
\tablecomments{The typical uncertainty for the dispersions is 5-10 km s$^{-1}$.}
\end{deluxetable}
\end{center}

\end{document}